\documentclass[aps,prb,twocolumn,nobibnotes,a4paper]{revtex4-1}
\usepackage[ansinew]{inputenc}
\usepackage{graphicx}
\usepackage{textcomp}

\newcommand{\bra}[1]{\langle #1 |}
\newcommand{\ket}[1]{|#1\rangle}

\begin{document}

\title{A quantum gate between a flying optical photon and a single trapped atom}
\author{Andreas~Reiserer}
\author{Norbert~Kalb}
\author{Gerhard~Rempe}
\email{gerhard.rempe@mpq.mpg.de}
\author{Stephan~Ritter}

\affiliation{Max-Planck-Institut f\"ur Quantenoptik, Hans-Kopfermann-Strasse 1, 85748 Garching, Germany}

\maketitle
\textbf{
The steady increase in control over individual quantum systems has backed the dream of a quantum technology that provides functionalities beyond any classical device. Two particularly promising applications have been explored during the past decade: First, photon-based quantum communication, which guarantees unbreakable encryption \cite{gisin_quantum_2002} but still has to be scaled to high rates over large distances. Second, quantum computation, which will fundamentally enhance computability \cite{ladd_quantum_2010} if it can be scaled to a large number of quantum bits. It was realized early on that a hybrid system of light and matter qubits \cite{duan_colloquium:_2010} could solve the scalability problem of both fields---that of communication via quantum repeaters \cite{briegel_quantum_1998}, that of computation via an optical interconnect between smaller quantum processors \cite{monroe_scaling_2013, awschalom_quantum_2013}. To this end, the development of a robust two-qubit gate that allows to link distant computational nodes is ``a pressing challenge'' \cite{awschalom_quantum_2013}.
Here we demonstrate such a quantum gate between the spin state of a single trapped atom and the polarization state of an optical photon contained in a faint laser pulse. The presented gate mechanism \cite{duan_scalable_2004} is deterministic, robust and expected to be applicable to almost any matter qubit. It is based on reflecting the photonic qubit from a cavity that provides strong light-matter coupling. To demonstrate its versatility, we use the quantum gate to create atom-photon, atom-photon-photon, and photon-photon entangled states from separable input states. We expect our experiment to break ground for various applications, including the generation of atomic \cite{cho_generation_2005} and photonic \cite{schon_sequential_2005, hu_deterministic_2008} cluster states, Schrödinger-cat states \cite{wang_engineering_2005}, deterministic photonic Bell-state measurements \cite{bonato_cnot_2010}, and quantum communication using a redundant quantum parity code \cite{munro_quantum_2012}.
}

\begin{figure}
\includegraphics[width=1.0\columnwidth]{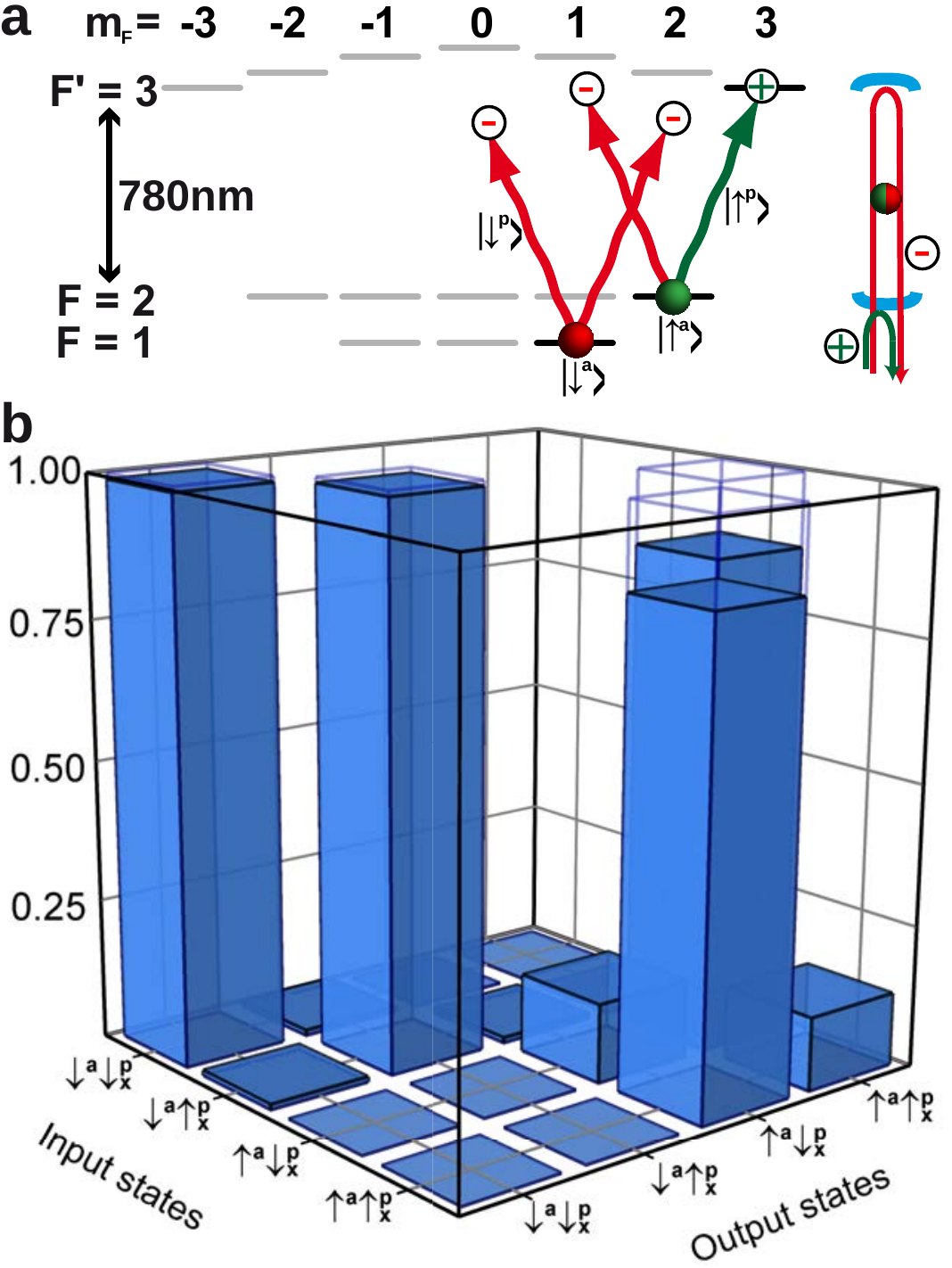}
\caption{\label{fig:Fig1}
\textbf{Atom-photon quantum gate.} \textbf{a}, Atomic level scheme on the D$_2$ line of $^{87}$Rb. The photonic qubit is defined in the basis of left- ($\ket{\downarrow^p}$) and right- ($\ket{\uparrow^p}$) circular polarization. The atomic qubit is encoded in the atomic $\ket{F,m_F}$ states $\ket{\downarrow^a}\equiv\ket{1,1}$ and $\ket{\uparrow^a}\equiv\ket{2,2}$. Here, $F$ denotes the atomic hyperfine state and $m_F$ its projection onto an external magnetic field. The cavity (blue semi-circles) is resonant with the a.c. Stark-shifted $\ket{2,2} \leftrightarrow \ket{3,3}$ transition on the D$_2$ line around 780\,nm. Upon reflection of a photon from the cavity, the combined atom-photon state $\ket{\uparrow^a \uparrow^p}$ (green, $\oplus$) acquires a phase shift of $\pi$ with respect to all other states (red, $\ominus$). \textbf{b}, Measured truth table. The bars represent the normalized probability of obtaining a certain output state for a complete orthogonal set of input states. Open blue bars indicate the action of an ideal controlled-NOT gate.
}
\end{figure}

Since their infancy, the fields of quantum communication and quantum computation have been largely independent. For communication \cite{gisin_quantum_2002}, optical photons are employed because they allow to transmit quantum states, such as time-bin or polarization qubits, over large distances using existing telecommunication fibre technology. Quantum computation \cite{ladd_quantum_2010}, on the other hand, is typically based on single spins, either in vacuum or in specific solid-state host materials. In addition to the long coherence times these spins can exhibit, they provide deterministic interaction mechanisms that facilitate local two-qubit quantum gates. Scalability would be offered by combining the specific advantages of both information carriers, spins and photons \cite{monroe_scaling_2013, awschalom_quantum_2013}. To implement the required interaction between the different types of qubits, a deterministic quantum gate between a photon and an atom has been proposed \cite{duan_scalable_2004}. Here, we demonstrate this quantum gate and its potential for quantum information processing with atoms and photons.

The employed mechanism is based on cavity quantum electrodynamics. When a photon interacts with a cavity containing a single, resonant emitter, it experiences a phase shift \cite{turchette_measurement_1995, fushman_controlled_2008} which depends on the coupling strength. In our experiment, the emitter is a single $^{87}$Rb atom, which is trapped at the centre of an overcoupled cavity. Full control over the position and motion of the atom \cite{reiserer_ground-state_2013} puts the system into the strong coupling regime (measured coupling constant $g=2\pi\cdot6.7$\,MHz, atomic dipole decay rate $\gamma=2\pi\cdot3$\,MHz, cavity field decay rate $\kappa=2\pi\cdot2.5$\,MHz). In this regime, the conditional phase shift induced on a reflected light field is $\pi$ \cite{reiserer_nondestructive_2013}, which is the prerequisite for the quantum gate presented in this work.

In contrast to the original proposal \cite{duan_scalable_2004}, our implementation does not require interferometric stability, as the a.c. Stark shift of a linearly polarized dipole trap is used to split the Zeeman states of the excited atomic state manifold (see Methods and the level scheme in Fig.\,\ref{fig:Fig1}a). Thus, the coupling is only strong when the atom is in state $\ket{\uparrow^a}$ and photons of right-circular polarization $\ket{\uparrow^p}$ are reflected (green arrow and sphere). For all other qubit combinations (red arrows and sphere), the coupling is negligible because any atomic transition is detuned (see Methods). Therefore, the reflection of a photon results in a conditional phase shift of $\pi$, i.e. a sign change, between the atomic and the photonic qubit:
\begin{eqnarray*}
\ket{\uparrow^a \uparrow^p}     & \rightarrow   & \ket{\uparrow^a \uparrow^p}\\
\ket{\downarrow^a \uparrow^p}   & \rightarrow - & \ket{\downarrow^a \uparrow^p}\\
\ket{\uparrow^a \downarrow^p}   & \rightarrow - & \ket{\uparrow^a \downarrow^p}\\
\ket{\downarrow^a \downarrow^p} & \rightarrow - & \ket{\downarrow^a \downarrow^p}
\end{eqnarray*}
This conditional phase shift allows to construct a universal quantum gate that can be transformed into any two-qubit gate using rotations of the individual qubits, which are implemented with wave plates for the photon and Raman transitions for the atom. With respect to the photonic basis states $\ket{\uparrow_x^p} \equiv \frac{1}{\sqrt{2}} (\ket{\uparrow^p} + \ket{\downarrow^p})$ and $\ket{\downarrow_x^p} \equiv \frac{1}{\sqrt{2}} (\ket{\uparrow^p} - \ket{\downarrow^p})$, the conditional phase shift represents an atom-photon controlled-NOT (CNOT) gate.

The action of the quantum CNOT gate is a flip of the photonic target qubit, controlled by the quantum state of the atom, similar to its classical analogue. A first step to characterize the gate is therefore to measure a classical truth table. To this end, the atomic state is prepared by optical pumping either into the uncoupled $F=1$ states, corresponding to $\ket{\downarrow^a}$, or into the coupled $\ket{\uparrow^a}$ state (see Methods). Subsequently, faint laser pulses of Gaussian temporal shape (average photon number $\bar{n}=0.3$, full width at half maximum (FWHM) 0.7\,\textmu s) in $\ket{\downarrow^p_x}$ or $\ket{\uparrow^p_x}$ are reflected from the cavity and measured with single-photon counting modules in a polarization-resolving setup. Then, the atomic state is measured within 3\,\textmu s using cavity-enhanced hyperfine-state detection (see Methods, Extended Data Fig.\,\ref{fig:StateDetection} and Ref.\,\cite{reiserer_nondestructive_2013}). The results are shown in Fig.\,\ref{fig:Fig1}b (see also Extended Data Table \ref{data:table}a), where the bars represent the normalized probabilities to detect a certain output state for each of the orthogonal input states.

The control and target qubits are expected to be unchanged when the control qubit is in the state $\ket{\downarrow^a}$, which is accomplished with a probability of 99\%. This number is limited by imperfections in the detection of the photon polarization and the atomic hyperfine state. When the control qubit is in $\ket{\uparrow^a}$, the expected flip of the photonic target qubit is observed with a probability of $86\%$. The statistical errors in the depicted data are negligible. However, we observe ambient-temperature-related drifts of about 2\% on a timescale of several hours. The flip probability is predominantly limited by two effects: First, by optical mode matching, because the transverse overlap between the free-space mode of the photon and the cavity mode is $92(3)\%$. Second, by the quality of preparing the state $\ket{\uparrow^a}$, which is successful with 96(1)\% probability. Finally, by the relative stability between the cavity resonance and the frequency of the impinging laser pulse, which is about 300\,kHz. None of these imperfections has a fundamental limit.

\begin{figure}
\includegraphics[width=1.0\columnwidth]{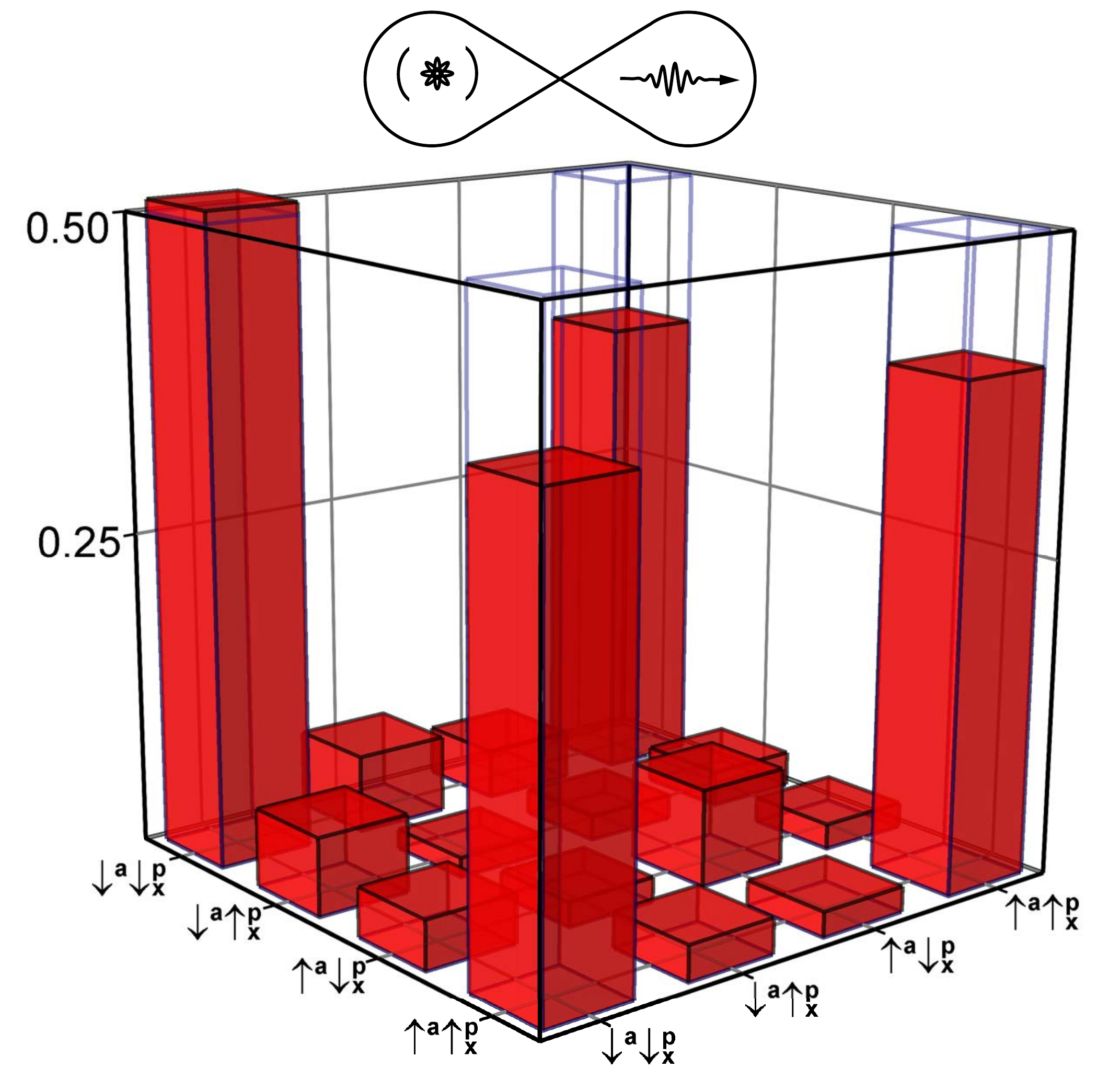}
\caption{\label{fig:TwoParticleEntanglement}
\textbf{Entangled atom-photon state generated via the gate operation.} The bars show the absolute value of the density-matrix elements. The fidelity with the maximally entangled $\ket{\Phi^+_{ap}}$ Bell state (open blue bars) is 80.7(0.5)\%. The icon at the top of the figure symbolizes atom-photon entanglement.
}
\end{figure}

The decisive feature that discriminates a quantum gate from a classical one is the generation of entangled states from separable input states. To characterize this property, faint laser pulses ($\bar{n}=0.07$, FWHM 0.7\,\textmu s) are reflected from the setup and the evaluation is post-selected on those cases where a single photon has subsequently been detected. The input state is $\ket{\downarrow^a_x \downarrow^p_x}$, such that the gate generates the maximally entangled $\ket{\Phi^+_{ap}}$ state:

$$\ket{\downarrow^a_x \downarrow^p_x} \rightarrow \ket{\Phi^+_{ap}} = \frac{1}{\sqrt{2}} (\ket{\uparrow^a \uparrow^p_x}+\ket{\downarrow^a \downarrow^p_x})$$

Both, the atomic and photonic qubit are measured in three orthogonal bases. This allows to reconstruct the density matrix $\rho_{ap}$ of the combined atom-photon state using quantum-state tomography and a maximum-likelihood estimation \cite{paris_quantum_2004}. The result is shown in Fig.\,\ref{fig:TwoParticleEntanglement} (see also Extended Data Table \ref{data:table}b). In accordance with the truth table measurement above, the density matrix is slightly asymmetric. While the value of $\ket{\downarrow^a \downarrow^p_x}\bra{\downarrow^a \downarrow^p_x}$ (left corner) is close to the ideal 0.5, the elements in the other corners are smaller. The fidelity with the expected $\ket{\Phi^+_{ap}}$ state is $F_{\Phi^+_{ap}}=\bra{\Phi^+_{ap}}\rho_{ap}\ket{\Phi^+_{ap}}=80.7(0.5)\%$, where the standard error has been determined with the Monte-Carlo technique \cite{paris_quantum_2004}. In the depicted measurement, the fidelity with a slightly rotated, maximally entangled state of the form \mbox{$\frac{1}{\sqrt{2}} (\ket{\uparrow^a \uparrow^p_x} + e^{-i\varphi} \ket{\downarrow^a \downarrow^p_x})$} can be higher, probably due to a small frequency offset between the cavity and the photon. We find a maximum value of $83.0\%$ for $\varphi = 0.11\pi$.

The major experimental imperfections that reduce the fidelity are: First, the mentioned frequency and mode mismatch between cavity and impinging photon (estimated reduction: 8(3)\%); second, the quality of our atomic state preparation, rotation and readout (reduction 5(1)\%; see Extended Data Fig.\,\ref{fig:Ramsey}); third, imperfections in the photonic state measurement (e.g. detector dark counts, imperfect beam splitters; reduction 2\%); finally, the small probability to have more than one photon in the impinging laser pulses (reduction 2\%). Again, none of these limitations is fundamental.

In principle, the gate mechanism presented in this work is deterministic. In our experimental implementation, the photon is not back-reflected from the coupled system $\ket{\uparrow^a \uparrow^p}$ with a probability of 34(2)\% and in the uncoupled cases with a probability of 30(2)\% \cite{reiserer_nondestructive_2013}. The small difference in reflectivity also contributes slightly $(<1\%)$ to the observed reduction in fidelity \cite{cho_generation_2005}. The achieved loss level nevertheless allows for scalable quantum computation \cite{duan_efficient_2005} and deterministic quantum state transfer \cite{enk_photonic_1998}. One would still observe nonclassical correlations without post-selection in case a perfect single-photon source and a perfect detector were used to characterize our device. Besides, we expect that it is possible to dramatically improve the achieved value in next-generation cavities with increased atom-cavity coupling strength \cite{Dayan_photon_2008, volz_measurement_2011, thompson_coupling_2013, oshea_fiber-optical_2013} and reduced losses.

\begin{figure}
\includegraphics[width=1.0\columnwidth]{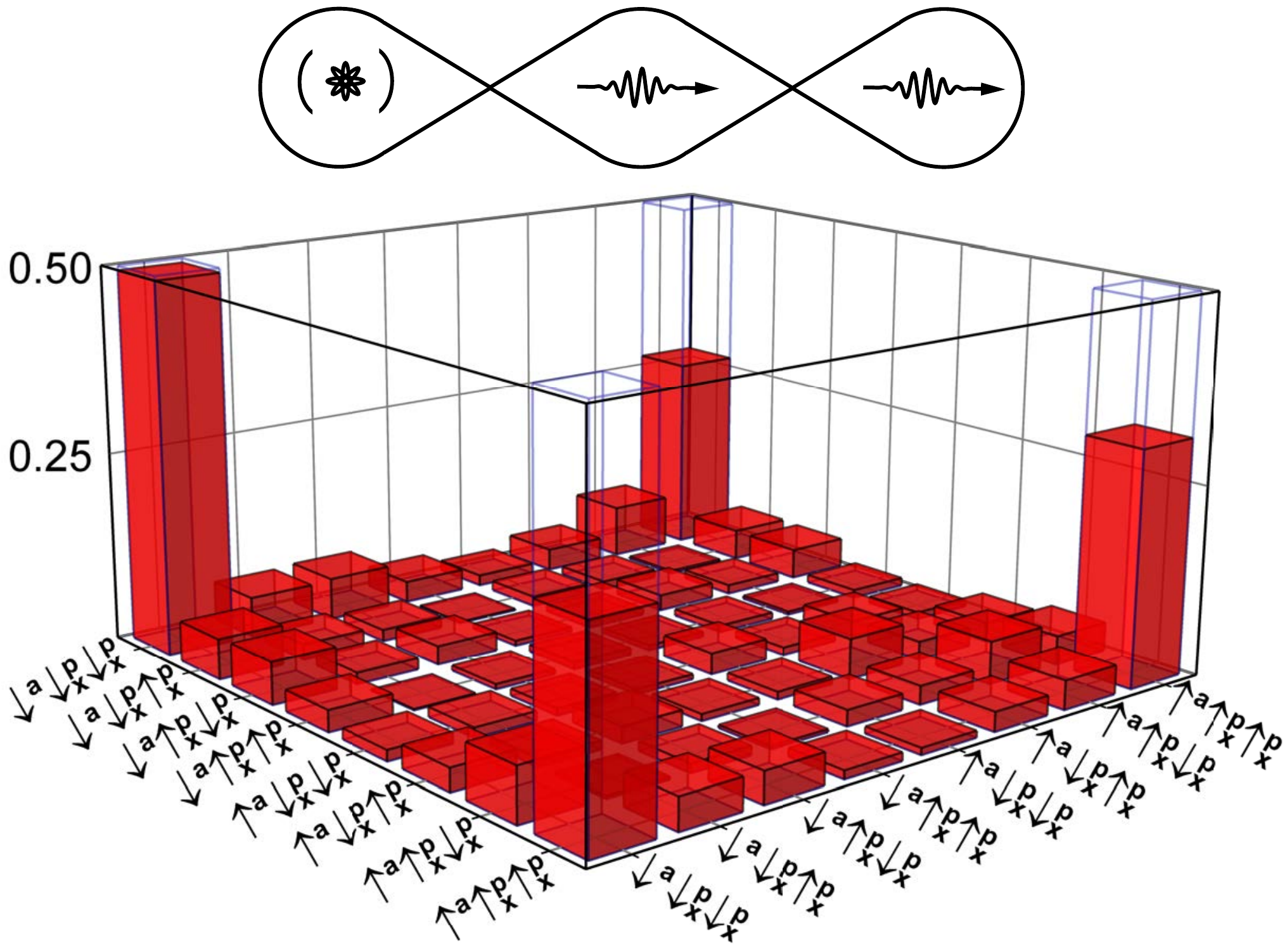}
\caption{\label{fig:ThreeParticleEntanglement}
\textbf{Entangled state between one atom and two photons.} This state is generated by reflecting two faint laser pulses from the cavity. The bars show the absolute value of the reconstructed density matrix elements. The fidelity with the maximally entangled state $\ket{\mathrm{GHZ}}$ is 61(2)\%. The matrix elements of $\ket{\mathrm{GHZ}}$ are depicted as open blue bars. The icon at the top of the figure symbolizes entanglement between an atom and two photons.
}
\end{figure}

The demonstrated quantum gate also allows to generate entangled cluster states that consist of the atom and several photons. To demonstrate this, the gate is applied to the photons contained in two sequentially impinging laser pulses (temporal distance 3\,\textmu s). Post-selecting events where one photon was detected in each of the input pulses, a maximally entangled Greenberger-Horne-Zeilinger (GHZ) state is expected:

$$\ket{\downarrow^a_x \downarrow^p_x \downarrow^p_x} \rightarrow \ket{\mathrm{GHZ}} = \frac{1}{\sqrt{2}} (\ket{\uparrow^a \uparrow^p_x \uparrow^p_x} - \ket{\downarrow^a \downarrow^p_x \downarrow^p_x})$$

The density matrix of the generated quantum state, again reconstructed using quantum state tomography and a maximum-likelihood estimation, is shown in Fig.\,\ref{fig:ThreeParticleEntanglement} (see also Extended Data Table \ref{data:table}c). The fidelity with the ideal state $\ket{\mathrm{GHZ}}$ (open blue bars in Fig.\,\ref{fig:ThreeParticleEntanglement}) is 61(2)\%, proving genuine three-particle (atom-photon-photon) entanglement. The reasons for a non-unity fidelity are analogous to the case of two particles. Again, we experimentally find a higher fidelity of 67\% with the slightly rotated GHZ state $\frac{1}{\sqrt{2}} (\ket{\uparrow^a \uparrow^p_x \uparrow^p_x} - e^{-i \varphi} \ket{\downarrow^a \downarrow^p_x \downarrow^p_x})$, with $\varphi=0.21\pi$.

Finally, we investigate whether the presented gate mechanism can mediate a photon-photon interaction for optical quantum computing \cite{duan_scalable_2004}. We employ a quantum eraser protocol \cite{roos_control_2004, hu_deterministic_2008} which should allow to create a maximally entangled state out of two separable input photons. To this end, the state $\ket{\mathrm{GHZ}}$ is generated as described above and a $\frac{\pi}{2}$ rotation is applied to the atom, which transforms the state to:

$$ \frac{1}{\sqrt{2}} \left[ \ket{\uparrow^a} (\ket{\uparrow^p_x \uparrow^p_x}-\ket{\downarrow^p_x \downarrow^p_x}) - \ket{\downarrow^a} (\ket{\uparrow^p_x \uparrow^p_x}+\ket{\downarrow^p_x \downarrow^p_x}) \right] $$

\begin{figure}
\includegraphics[width=1.0\columnwidth]{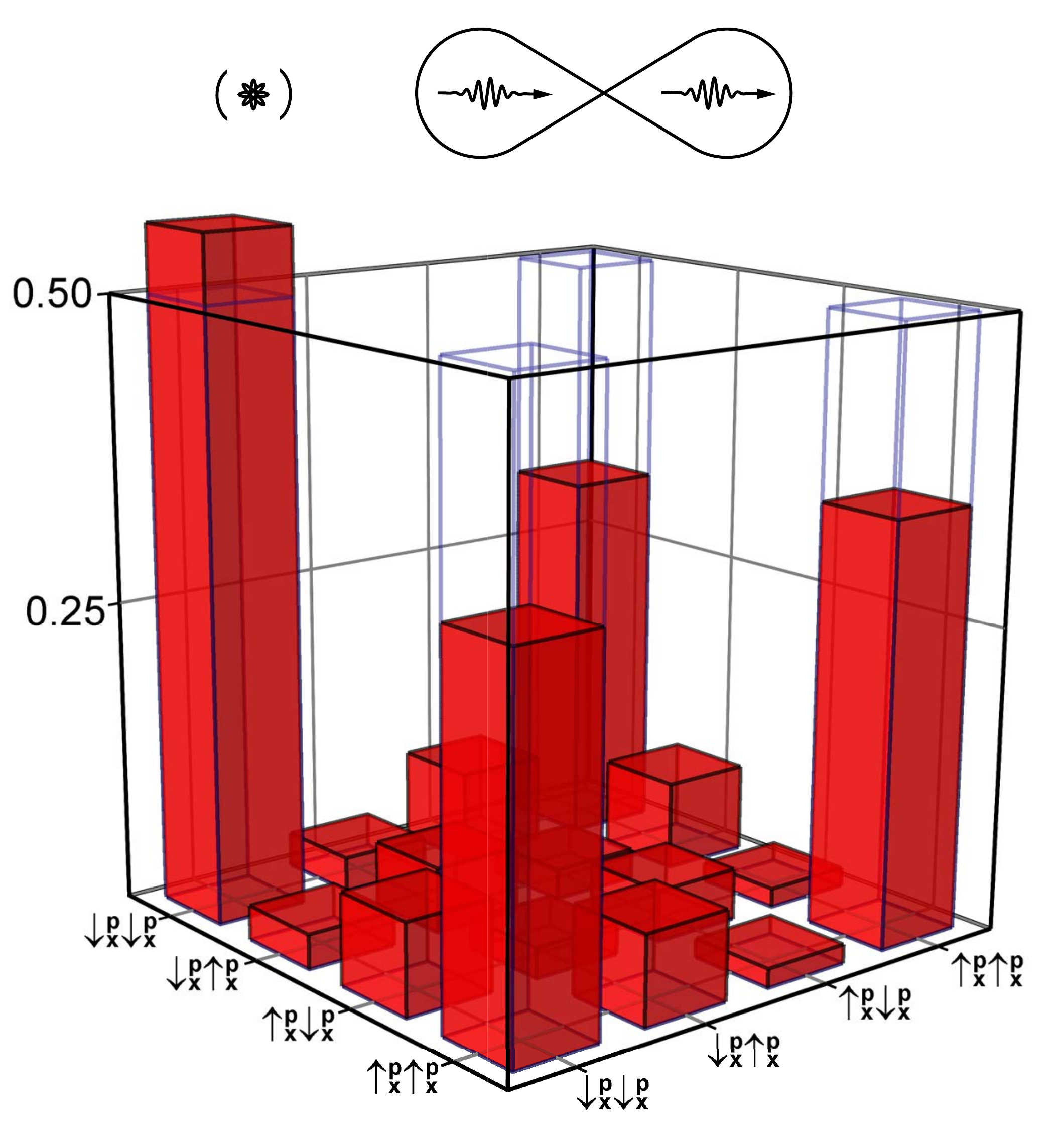}
\caption{\label{fig:TwoParticle_PhotonPhoton}
\textbf{Entangled photon-photon state generated via consecutive interaction with the atom.} The bars show the absolute value of the density matrix elements. The fidelity with the maximally entangled $\ket{\Phi^+_{pp}}$ Bell state (open blue bars) is 67(2)\%. The icon at the top of the figure symbolizes photon-photon entanglement.
}
\end{figure}

Subsequent measurement of the atomic state disentangles the atom, which results in a maximally entangled two-photon state: If the atom is found in $\ket{\downarrow^a}$ ($\ket{\uparrow^a}$), the resulting state is $\ket{\Phi^+_{pp}}$ ($\ket{\Phi^-_{pp}}$), respectively. In the experiment, the two-photon density matrices are again reconstructed with the maximum-likelihood technique (see Fig.\,\ref{fig:TwoParticle_PhotonPhoton} and Extended Data Table \ref{data:table}d and \ref{data:table}e). This gives a fidelity with the expected Bell states of $67(2)\%$ ($64(2)\%$) for the $\ket{\Phi^+_{pp}}$ ($\ket{\Phi^-_{pp}}$) state. The achieved values proof photon-photon entanglement. Their small difference can be explained by the fact that a detection of the atom in $F=1$ selects only those events where it has initially been prepared in the correct state $\ket{\uparrow^a}$, rather than in another state of the $F=2$ hyperfine manifold. Again, we find a higher fidelity of maximally $76\%$ with a rotated $\ket{\Phi^+_{pp}}$ state with $\varphi = 0.25\pi$.

The above measurements demonstrate the versatility of the presented gate mechanism and its ability to mediate a photon-photon interaction. To this end, intermediate storage of the two photons during the time required to rotate and read out the atomic state (about 3\,\textmu s) is required, which can be implemented with an optical fibre of less than one kilometre length. Conditioned on the state of the atom, the polarization of the photons then has to be rotated, e.g. using an electro-optical modulator. As an alternative to the eraser-scheme employed in this work, the first photon could be reflected from the cavity a second time \cite{duan_scalable_2004}.

In addition to the applications mentioned above \cite{monroe_scaling_2013, awschalom_quantum_2013, duan_scalable_2004, cho_generation_2005, schon_sequential_2005, hu_deterministic_2008, wang_engineering_2005, bonato_cnot_2010, munro_quantum_2012}, the presented gate mechanism opens up perspectives for numerous quantum optics experiments. First, it can be applied to perform a quantum-non-demolition measurement of the polarization of a single reflected photon by measuring the state of the atom. Vice versa, it can be used to measure the atomic state without energy exchange \cite{volz_measurement_2011} by measuring the polarization of a reflected photon. Besides, a quantum gate between several atoms in the same or even in remote cavities \cite{xiao_realizing_2004, duan_robust_2005} can be directly implemented, which also facilitates universal quantum computation in a decoherence-free subspace \cite{xue_universal_2006}. Finally, the proposed deterministic optical Bell-state measurement \cite{bonato_cnot_2010}would dramatically increase the efficiency of teleportation between remote atoms \cite{nolleke_efficient_2013} and therefore the prospects for the implementation of a quantum repeater \cite{briegel_quantum_1998} and a quantum network \cite{duan_colloquium:_2010, Ritter_elementary_2012} on a global scale.

\begin{acknowledgments}
This work was supported by the European Union (Collaborative Project SIQS) and by the Bundesministerium f\"ur Bildung und Forschung via IKT 2020 (QK\_QuOReP).
\end{acknowledgments}

\renewcommand{\figurename}{\textbf{Extended Data Figure}}
\renewcommand{\thefigure}{\arabic{figure}}
\setcounter{figure}{0}

\section*{Methods}
\subsection*{Experimental setup}
In the experimental setup, single $^{87}$Rb atoms are loaded from a magneto-optical trap into a three-dimensional optical lattice inside a Fabry-Perot cavity. The cavity is overcoupled, i.e. the coupling mirror has a transmission (95\,ppm) which is large compared to the transmission of the high-reflector and the scattering and absorption losses (8\,ppm total). The geometry of the trap and the employed cooling mechanisms are described in detail in Ref.\,\cite{reiserer_ground-state_2013}. The lattice consists of three retro-reflected laser beams, one red detuned (1064\,nm) and two blue detuned (770\,nm) from the atomic transitions at 780\,nm (D$_2$ line) and 795\,nm (D$_1$ line). The use of high intensities leads to trap frequencies of several hundred kHz, which facilitates fast cooling to low temperatures using intra-cavity Sisyphus cooling. In each experimental cycle, a cooling interval of 0.8\,ms is applied, which allows for atom trapping times of many seconds. In contrast to Ref.\,\cite{reiserer_ground-state_2013}, ground-state cooling is not applied in this work.

\subsection*{Light shift}
For the working principle of the gate, the a.c. Stark shift of the atomic levels is important, which is schematically depicted in Fig.\,\ref{fig:Fig1}a. In this context, the blue detuned trap light has a negligible influence, because the atom is trapped at a node of the standing-wave light field. The red-detuned light, however, considerably shifts the frequency of the atomic transitions, depending on the polarization of the trap laser. We employ $\pi$-polarized light, i.e. the electric field vector is oriented parallel to the quantization axis, which coincides with the cavity axis and the direction of an externally applied magnetic field of about 0.5\,G. In this configuration, the a.c. Stark shift is identical for all Zeeman states in the ground-state manifolds with $F=1$ and $F=2$. The excited state $F'=3$, however, experiences a Zeeman-state dependent shift. Its value has been measured for the atomic transition $\ket{2,2} \leftrightarrow \ket{3,3}$ to be 0.10\,GHz, where 0.05\,GHz stem from the shift of the ground state $\ket{2,2}$. The shifts of the other states in the $F'=3$ manifold can be calculated by summing over all relevant atomic levels considering the individual transition strengths. This leads to the following level shifts: $\ket{3,0}$: 0.16\,GHz, $\ket{3,\pm 1}$: 0.15\,GHz, $\ket{3,\pm2}$: 0.10\,GHz, $\ket{3,\pm3}$: 0.05\,GHz.

In the context of the gate mechanism, the impinging photon is on resonance with the transition $\ket{2,2}\leftrightarrow \ket{3,3}$. The transition $\ket{2,2}\leftrightarrow \ket{3,1}$ is thus detuned by 0.1\,GHz, while all transitions from the F=1 state are detuned by about 7\,GHz. Therefore, only the atom in state $\ket{\uparrow^a}$ and the photon in $\ket{\uparrow^p}$ are strongly coupled.

\subsection*{Atomic state preparation}
To prepare the atom in the state $\ket{2,2}$, a 140-\textmu s-long interval of optical pumping is used, where circularly polarized light is applied on resonance along the cavity axis and an additional repumping laser depletes the states with $F=1$. Once the atom is pumped to the desired state, the transmission of the pump light is strongly reduced due to strong coupling. Monitoring the cavity output with single-photon counting modules thus allows to preselect those experimental runs in which the atom has been pumped to the right state with high probability \cite{reiserer_nondestructive_2013}. The experimental results presented in this work were obtained by selecting those cases where one or zero photons were detected in the last 10 \textmu s of the optical pumping interval, which is about half of all attempts.

\subsection*{Atomic state detection}
\begin{figure}
\includegraphics[width=1.0\columnwidth]{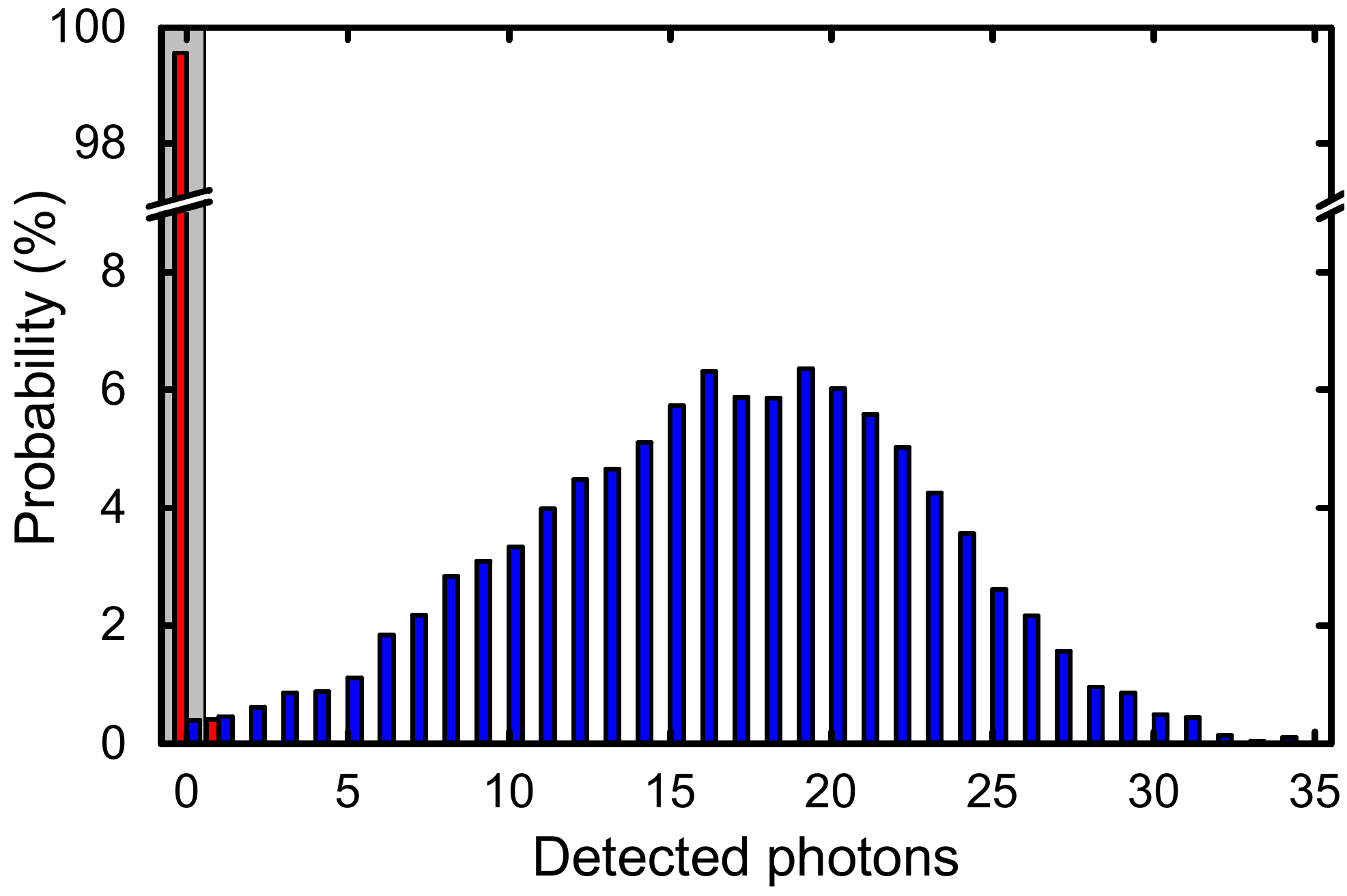}
\caption{\label{fig:StateDetection}
\textbf{Detection of the atomic state.} The atom is prepared either in the resonant $\ket{2,2}$ state (blue) or in the detuned $F=1$ state (red) and a resonant laser is irradiated for 3\,\textmu s from the side of the cavity. The number of photons detected in the cavity output allows to distinguish the two cases with a fidelity of 99.65\%.
}
\end{figure}

To detect the atomic hyperfine state, cavity-enhanced fluorescence state detection is employed. To this end, a laser resonant with the cavity is applied from the side \cite{reiserer_ground-state_2013}. If the atom is in the state $F=2$, it scatters many photons into the cavity (enhanced by the Purcell effect), while there is no scattering if the atom is in the state $F=1$ due to the large atom-laser detuning of 7\,GHz. The described technique only detects the hyperfine state and is not sensitive to the atomic Zeeman state. Compared to previous experiments \cite{reiserer_nondestructive_2013}, the time required to detect the atomic state could be reduced from 25\,\textmu s to 3\,\textmu s by increasing the power of the applied laser beam. Further reduction is possible, however at the price of a reduced atom-trapping time. A histogram of the detected photon number per run is depicted in Extended Data Fig.\,\ref{fig:StateDetection}. When the atom is prepared in a state with $F=1$ (red), no photons are detected in 99.7\% of all cases. When the atom is prepared in $\ket{2,2}$ (blue), one or more photons are detected in 99.6\% of the runs. Thus, when setting the threshold between 0 and 1 detected photons, the state detection fidelity is 99.65\%.

\subsection*{Atomic state rotation}
\begin{figure}
\includegraphics[width=1.0\columnwidth]{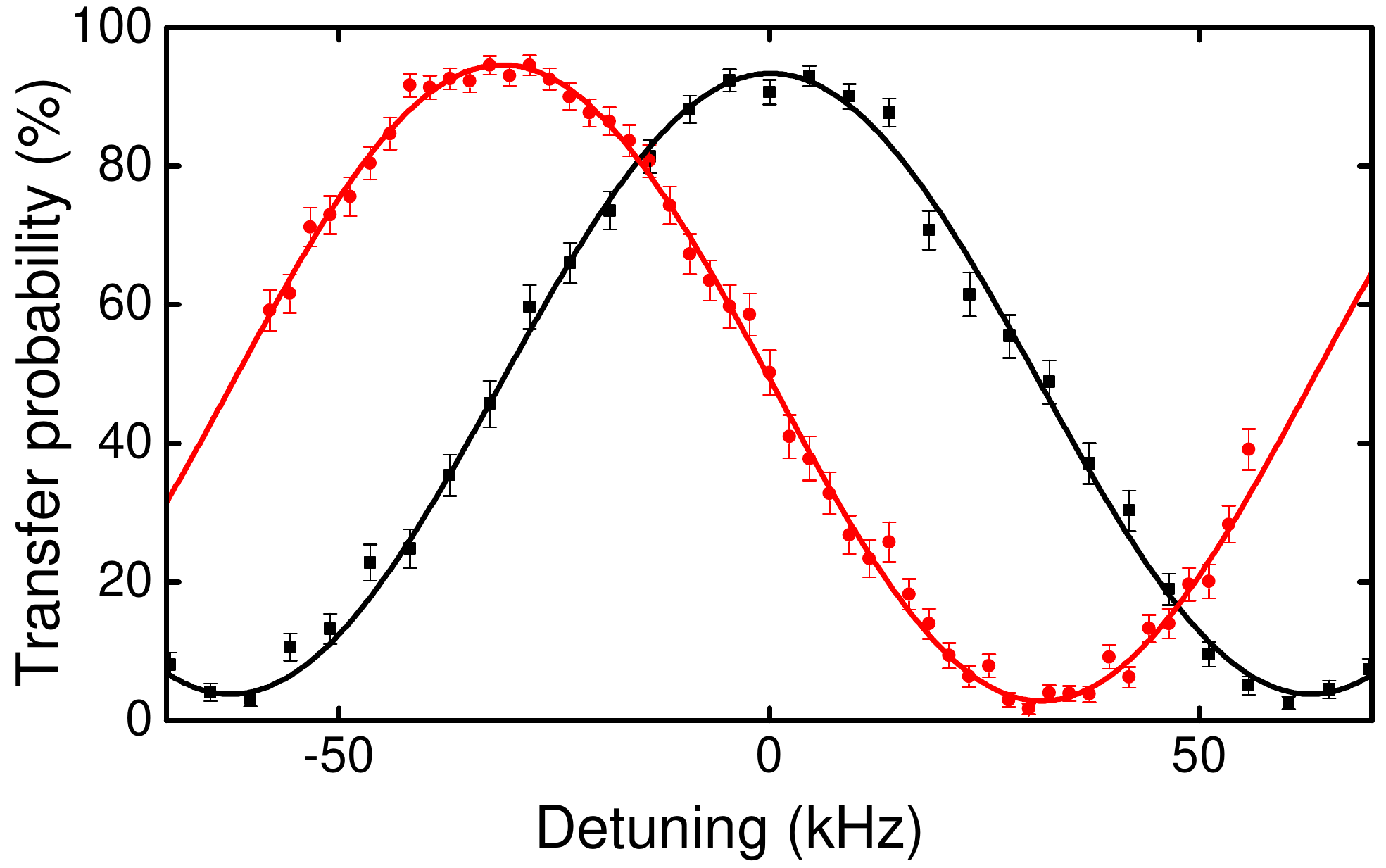}
\caption{\label{fig:Ramsey}
\textbf{Ramsey spectroscopy.} The atom is prepared in the state $\ket{2,2}$ and two $\frac{\pi}{2}$ Raman pulses are applied with a temporal distance of 7.5\,\textmu s. Scanning the Raman laser detuning, a sinusoidal oscillation is observed. When the second pulse is applied with a phase shift of $\frac{\pi}{2}$ (red), the curve is shifted by a quarter of a period with respect to the case without phase shift (black). From the amplitude of the sinusoidal fit curves, we deduce that the atomic state preparation, rotation and readout works as intended in 95(1)\% of the experiments.
}
\end{figure}

In order to rotate the atomic state, we employ a pair of co-propagating Raman laser beams with orthogonal polarization, applied from the side of the cavity with a detuning of $-0.15$\,THz to the D$_1$ line. A magnetic field applied along the cavity axis splits the atomic Zeeman states by 0.3\,MHz, which allows to spectrally address individual transitions. To investigate the quality of the combined atomic-state preparation, rotation and readout process, Ramsey spectroscopy is performed. To this end, the atom is prepared in the state $\ket{2,2}$ and two $\frac{\pi}{2}$ Raman pulses are applied (duration: 1.7\,\textmu s; temporal distance: 7.5\,\textmu s). The result of a subsequent measurement of the atomic state is shown in Extended Data Fig.\,\ref{fig:Ramsey}. At zero detuning, this sequence would ideally result in a transfer probability of 100\% when the two pulses are applied with the same phase (black). Experimentally, we observe 95(1)\%. Scanning the laser frequency over a few ten kHz, a sinusoidal oscillation is observed, which, as expected, shifts by a quarter of a period when the second Ramsey pulse is applied with a phase difference of $\frac{\pi}{2}$ (red). From the difference between the maximum and minimum values of the observed curve, 90(2)\%, we conclude that the atomic state preparation, rotation and readout process works as intended in 95(1)\% of the experiments, which includes dephasing during the 7.5\,\textmu s between the Raman pulses.

\newpage
\renewcommand{\figurename}{\textbf{Extended Data Table}}
\renewcommand{\thefigure}{\arabic{figure}}
\setcounter{figure}{0}

\begin{figure*}
\includegraphics[width=\textwidth]{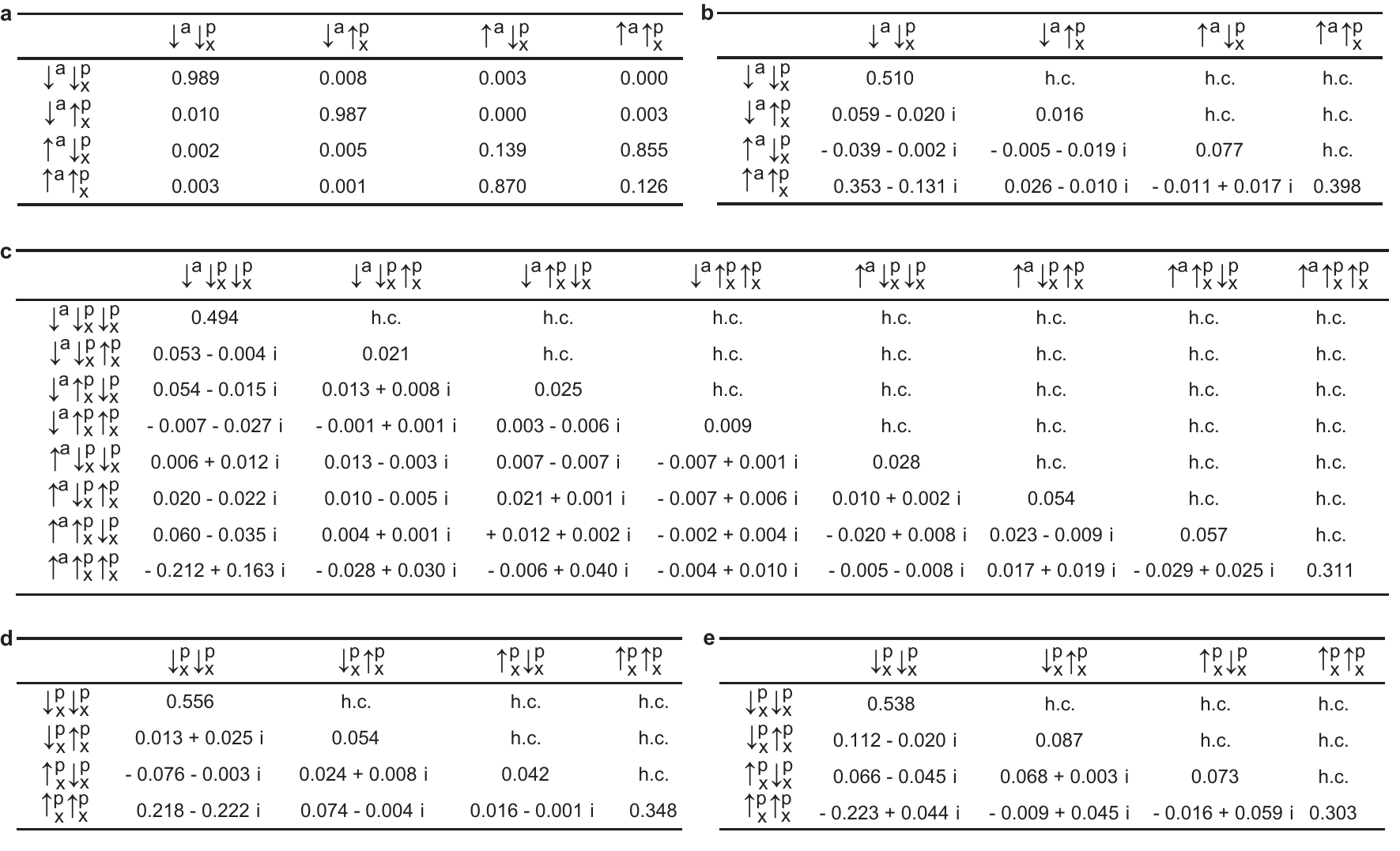}
\caption{\label{data:table}
\textbf{Numerical values of the truth table and density matrices.} h.c. denotes the Hermitian conjugate. \textbf{a}, Data of the truth table measurement depicted in Fig.\,\ref{fig:Fig1}. \textbf{b}, Atom-photon density matrix. The absolute values of the elements are depicted in Fig.\,\ref{fig:TwoParticleEntanglement}. \textbf{c}, Atom-photon-photon density matrix. The absolute value of the elements are depicted in Fig.\,\ref{fig:ThreeParticleEntanglement}. \textbf{d}, Photon-photon density matrix, postselected on the detection of the atomic $\ket{\downarrow^a}$ state. The absolute value of the elements are depicted in Fig.\,\ref{fig:TwoParticle_PhotonPhoton}. \textbf{e}, Photon-photon density matrix, postselected on the detection of the atomic $\ket{\uparrow^a}$ state.
}
\end{figure*}


\begin{thebibliography}{10}
\expandafter\ifx\csname url\endcsname\relax
  \def\url#1{\texttt{#1}}\fi
\expandafter\ifx\csname urlprefix\endcsname\relax\def\urlprefix{URL }\fi
\providecommand{\bibinfo}[2]{#2}
\providecommand{\eprint}[2][]{\url{#2}}

\bibitem{gisin_quantum_2002}
\bibinfo{author}{Gisin, N.}, \bibinfo{author}{Ribordy, G.},
  \bibinfo{author}{Tittel, W.} \& \bibinfo{author}{Zbinden, H.}
\newblock \bibinfo{title}{Quantum cryptography}.
\newblock \emph{\bibinfo{journal}{Rev. Mod. Phys.}}
  \textbf{\bibinfo{volume}{74}}, \bibinfo{pages}{145--195}
  (\bibinfo{year}{2002}).

\bibitem{ladd_quantum_2010}
\bibinfo{author}{Ladd, T.~D.} \emph{et~al.}
\newblock \bibinfo{title}{Quantum computers}.
\newblock \emph{\bibinfo{journal}{Nature}} \textbf{\bibinfo{volume}{464}},
  \bibinfo{pages}{45--53} (\bibinfo{year}{2010}).

\bibitem{duan_colloquium:_2010}
\bibinfo{author}{Duan, L.-M.} \& \bibinfo{author}{Monroe, C.}
\newblock \bibinfo{title}{Colloquium: Quantum networks with trapped ions}.
\newblock \emph{\bibinfo{journal}{Rev. Mod. Phys.}}
  \textbf{\bibinfo{volume}{82}}, \bibinfo{pages}{1209--1224}
  (\bibinfo{year}{2010}).

\bibitem{briegel_quantum_1998}
\bibinfo{author}{Briegel, H.-J.}, \bibinfo{author}{D{\"u}r, W.},
  \bibinfo{author}{Cirac, J.~I.} \& \bibinfo{author}{Zoller, P.}
\newblock \bibinfo{title}{Quantum repeaters: The role of imperfect local
  operations in quantum communication}.
\newblock \emph{\bibinfo{journal}{Phys. Rev. Lett.}}
  \textbf{\bibinfo{volume}{81}}, \bibinfo{pages}{5932--5935}
  (\bibinfo{year}{1998}).

\bibitem{monroe_scaling_2013}
\bibinfo{author}{Monroe, C.} \& \bibinfo{author}{Kim, J.}
\newblock \bibinfo{title}{Scaling the ion trap quantum processor}.
\newblock \emph{\bibinfo{journal}{Science}} \textbf{\bibinfo{volume}{339}},
  \bibinfo{pages}{1164--1169} (\bibinfo{year}{2013}).

\bibitem{awschalom_quantum_2013}
\bibinfo{author}{Awschalom, D.~D.}, \bibinfo{author}{Bassett, L.~C.},
  \bibinfo{author}{Dzurak, A.~S.}, \bibinfo{author}{Hu, E.~L.} \&
  \bibinfo{author}{Petta, J.~R.}
\newblock \bibinfo{title}{Quantum spintronics: Engineering and manipulating
  atom-like spins in semiconductors}.
\newblock \emph{\bibinfo{journal}{Science}} \textbf{\bibinfo{volume}{339}},
  \bibinfo{pages}{1174--1179} (\bibinfo{year}{2013}).

\bibitem{duan_scalable_2004}
\bibinfo{author}{Duan, L.-M.} \& \bibinfo{author}{Kimble, H.~J.}
\newblock \bibinfo{title}{Scalable photonic quantum computation through
  cavity-assisted interactions}.
\newblock \emph{\bibinfo{journal}{Phys. Rev. Lett.}}
  \textbf{\bibinfo{volume}{92}}, \bibinfo{pages}{127902}
  (\bibinfo{year}{2004}).

\bibitem{cho_generation_2005}
\bibinfo{author}{Cho, J.} \& \bibinfo{author}{Lee, H.-W.}
\newblock \bibinfo{title}{Generation of atomic cluster states through the
  cavity input-output process}.
\newblock \emph{\bibinfo{journal}{Phys. Rev. Lett.}}
  \textbf{\bibinfo{volume}{95}}, \bibinfo{pages}{160501}
  (\bibinfo{year}{2005}).

\bibitem{schon_sequential_2005}
\bibinfo{author}{Sch{\"o}n, C.}, \bibinfo{author}{Solano, E.},
  \bibinfo{author}{Verstraete, F.}, \bibinfo{author}{Cirac, J.~I.} \&
  \bibinfo{author}{Wolf, M.~M.}
\newblock \bibinfo{title}{Sequential generation of entangled multiqubit
  states}.
\newblock \emph{\bibinfo{journal}{Phys. Rev. Lett.}}
  \textbf{\bibinfo{volume}{95}}, \bibinfo{pages}{110503}
  (\bibinfo{year}{2005}).

\bibitem{hu_deterministic_2008}
\bibinfo{author}{Hu, C.~Y.}, \bibinfo{author}{Munro, W.~J.} \&
  \bibinfo{author}{Rarity, J.~G.}
\newblock \bibinfo{title}{Deterministic photon entangler using a charged
  quantum dot inside a microcavity}.
\newblock \emph{\bibinfo{journal}{Phys. Rev. B}} \textbf{\bibinfo{volume}{78}},
  \bibinfo{pages}{125318} (\bibinfo{year}{2008}).

\bibitem{wang_engineering_2005}
\bibinfo{author}{Wang, B.} \& \bibinfo{author}{Duan, L.-M.}
\newblock \bibinfo{title}{Engineering superpositions of coherent states in
  coherent optical pulses through cavity-assisted interaction}.
\newblock \emph{\bibinfo{journal}{Phys. Rev. A}} \textbf{\bibinfo{volume}{72}},
  \bibinfo{pages}{022320} (\bibinfo{year}{2005}).

\bibitem{bonato_cnot_2010}
\bibinfo{author}{Bonato, C.} \emph{et~al.}
\newblock \bibinfo{title}{{CNOT} and {B}ell-state analysis in the weak-coupling
  cavity {QED} regime}.
\newblock \emph{\bibinfo{journal}{Phys. Rev. Lett.}}
  \textbf{\bibinfo{volume}{104}}, \bibinfo{pages}{160503}
  (\bibinfo{year}{2010}).

\bibitem{munro_quantum_2012}
\bibinfo{author}{Munro, W.~J.}, \bibinfo{author}{Stephens, A.~M.},
  \bibinfo{author}{Devitt, S.~J.}, \bibinfo{author}{Harrison, K.~A.} \&
  \bibinfo{author}{Nemoto, K.}
\newblock \bibinfo{title}{Quantum communication without the necessity of
  quantum memories}.
\newblock \emph{\bibinfo{journal}{Nature Photon.}}
  \textbf{\bibinfo{volume}{6}}, \bibinfo{pages}{777--781}
  (\bibinfo{year}{2012}).

\bibitem{turchette_measurement_1995}
\bibinfo{author}{Turchette, Q.~A.}, \bibinfo{author}{Hood, C.~J.},
  \bibinfo{author}{Lange, W.}, \bibinfo{author}{Mabuchi, H.} \&
  \bibinfo{author}{Kimble, H.~J.}
\newblock \bibinfo{title}{Measurement of conditional phase shifts for quantum
  logic}.
\newblock \emph{\bibinfo{journal}{Phys. Rev. Lett.}}
  \textbf{\bibinfo{volume}{75}}, \bibinfo{pages}{4710--4713}
  (\bibinfo{year}{1995}).

\bibitem{fushman_controlled_2008}
\bibinfo{author}{Fushman, I.} \emph{et~al.}
\newblock \bibinfo{title}{Controlled phase shifts with a single quantum dot}.
\newblock \emph{\bibinfo{journal}{Science}} \textbf{\bibinfo{volume}{320}},
  \bibinfo{pages}{769--772} (\bibinfo{year}{2008}).

\bibitem{reiserer_ground-state_2013}
\bibinfo{author}{Reiserer, A.}, \bibinfo{author}{N{\"o}lleke, C.},
  \bibinfo{author}{Ritter, S.} \& \bibinfo{author}{Rempe, G.}
\newblock \bibinfo{title}{Ground-state cooling of a single atom at the center
  of an optical cavity}.
\newblock \emph{\bibinfo{journal}{Phys. Rev. Lett.}}
  \textbf{\bibinfo{volume}{110}}, \bibinfo{pages}{223003}
  (\bibinfo{year}{2013}).

\bibitem{reiserer_nondestructive_2013}
\bibinfo{author}{Reiserer, A.}, \bibinfo{author}{Ritter, S.} \&
  \bibinfo{author}{Rempe, G.}
\newblock \bibinfo{title}{Nondestructive detection of an optical photon}.
\newblock \emph{\bibinfo{journal}{Science}} \textbf{\bibinfo{volume}{342}},
  \bibinfo{pages}{1349--1351} (\bibinfo{year}{2013}).

\bibitem{paris_quantum_2004}
\bibinfo{author}{Paris, M.} \& \bibinfo{author}{\v{R}eh\'{a}\v{c}ek, J.}
\newblock \emph{\bibinfo{title}{Quantum State Estimation}}
  (\bibinfo{publisher}{Springer}, \bibinfo{year}{2004}).

\bibitem{duan_efficient_2005}
\bibinfo{author}{Duan, L.-M.} \& \bibinfo{author}{Raussendorf, R.}
\newblock \bibinfo{title}{Efficient quantum computation with probabilistic
  quantum gates}.
\newblock \emph{\bibinfo{journal}{Phys. Rev. Lett.}}
  \textbf{\bibinfo{volume}{95}}, \bibinfo{pages}{080503}
  (\bibinfo{year}{2005}).

\bibitem{enk_photonic_1998}
\bibinfo{author}{van Enk, S.~J.}, \bibinfo{author}{Cirac, J.~I.} \&
  \bibinfo{author}{Zoller, P.}
\newblock \bibinfo{title}{Photonic channels for quantum communication}.
\newblock \emph{\bibinfo{journal}{Science}} \textbf{\bibinfo{volume}{279}},
  \bibinfo{pages}{205--208} (\bibinfo{year}{1998}).

\bibitem{Dayan_photon_2008}
\bibinfo{author}{Dayan, B.} \emph{et~al.}
\newblock \bibinfo{title}{A photon turnstile dynamically regulated by one
  atom}.
\newblock \emph{\bibinfo{journal}{Science}} \textbf{\bibinfo{volume}{319}},
  \bibinfo{pages}{1062--1065} (\bibinfo{year}{2008}).

\bibitem{volz_measurement_2011}
\bibinfo{author}{Volz, J.}, \bibinfo{author}{Gehr, R.},
  \bibinfo{author}{Dubois, G.}, \bibinfo{author}{Est\`{e}ve, J.} \&
  \bibinfo{author}{Reichel, J.}
\newblock \bibinfo{title}{Measurement of the internal state of a single atom
  without energy exchange}.
\newblock \emph{\bibinfo{journal}{Nature}} \textbf{\bibinfo{volume}{475}},
  \bibinfo{pages}{210--213} (\bibinfo{year}{2011}).

\bibitem{thompson_coupling_2013}
\bibinfo{author}{Thompson, J.~D.} \emph{et~al.}
\newblock \bibinfo{title}{Coupling a single trapped atom to a nanoscale optical
  cavity}.
\newblock \emph{\bibinfo{journal}{Science}} \textbf{\bibinfo{volume}{340}},
  \bibinfo{pages}{1202--1205} (\bibinfo{year}{2013}).

\bibitem{oshea_fiber-optical_2013}
\bibinfo{author}{O'Shea, D.}, \bibinfo{author}{Junge, C.},
  \bibinfo{author}{Volz, J.} \& \bibinfo{author}{Rauschenbeutel, A.}
\newblock \bibinfo{title}{Fiber-optical switch controlled by a single atom}.
\newblock \emph{\bibinfo{journal}{Phys. Rev. Lett.}}
  \textbf{\bibinfo{volume}{111}}, \bibinfo{pages}{193601}
  (\bibinfo{year}{2013}).

\bibitem{roos_control_2004}
\bibinfo{author}{Roos, C.~F.} \emph{et~al.}
\newblock \bibinfo{title}{Control and measurement of three-qubit entangled
  states}.
\newblock \emph{\bibinfo{journal}{Science}} \textbf{\bibinfo{volume}{304}},
  \bibinfo{pages}{1478--1480} (\bibinfo{year}{2004}).

\bibitem{xiao_realizing_2004}
\bibinfo{author}{Xiao, Y.-F.} \emph{et~al.}
\newblock \bibinfo{title}{Realizing quantum controlled phase flip through
  cavity {QED}}.
\newblock \emph{\bibinfo{journal}{Phys. Rev. A}} \textbf{\bibinfo{volume}{70}},
  \bibinfo{pages}{042314} (\bibinfo{year}{2004}).

\bibitem{duan_robust_2005}
\bibinfo{author}{Duan, L.-M.}, \bibinfo{author}{Wang, B.} \&
  \bibinfo{author}{Kimble, H.~J.}
\newblock \bibinfo{title}{Robust quantum gates on neutral atoms with
  cavity-assisted photon scattering}.
\newblock \emph{\bibinfo{journal}{Phys. Rev. A}} \textbf{\bibinfo{volume}{72}},
  \bibinfo{pages}{032333} (\bibinfo{year}{2005}).

\bibitem{xue_universal_2006}
\bibinfo{author}{Xue, P.} \& \bibinfo{author}{Xiao, Y.-F.}
\newblock \bibinfo{title}{Universal quantum computation in decoherence-free
  subspace with neutral atoms}.
\newblock \emph{\bibinfo{journal}{Phys. Rev. Lett.}}
  \textbf{\bibinfo{volume}{97}}, \bibinfo{pages}{140501}
  (\bibinfo{year}{2006}).

\bibitem{nolleke_efficient_2013}
\bibinfo{author}{N{\"o}lleke, C.} \emph{et~al.}
\newblock \bibinfo{title}{Efficient teleportation between remote single-atom
  quantum memories}.
\newblock \emph{\bibinfo{journal}{Phys. Rev. Lett.}}
  \textbf{\bibinfo{volume}{110}}, \bibinfo{pages}{140403}
  (\bibinfo{year}{2013}).

\bibitem{Ritter_elementary_2012}
\bibinfo{author}{Ritter, S.} \emph{et~al.}
\newblock \bibinfo{title}{An elementary quantum network of single atoms in
  optical cavities}.
\newblock \emph{\bibinfo{journal}{Nature}} \textbf{\bibinfo{volume}{484}},
  \bibinfo{pages}{195--200} (\bibinfo{year}{2012}).

\end{thebibliography}
\end{document}